# The Role of Workers in AI Ethics and Governance

By Nataliya Nedzhvetskaya and JS Tan

**Abstract** (250 words)

While the role of states, corporations, and international organizations in AI governance has been extensively theorized, the role of workers has received comparatively little attention. This chapter looks at the role that workers play in identifying and mitigating harms from AI technologies. Harms are the causally assessed "impacts" of technologies. They arise despite technical reliability and are not a result of technical negligence but rather of normative uncertainty around questions of safety and fairness in complex social systems. There is high consensus in the AI ethics community on the benefits of reducing harms but less consensus on mechanisms for determining or addressing harms. This lack of consensus has resulted in a number of collective actions by workers protesting how harms are identified and addressed in their workplace. We theorize the role of workers within AI governance and construct a model of harm reporting processes in AI workplaces. The harm reporting process involves three steps: identification, the governance decision, and the response. Workers draw upon three types of claims to argue for jurisdiction over questions of AI governance: subjection, control over the product of one's labor, and proximate knowledge of systems. Examining the past decade of AI-related worker activism allows us to understand how different types of workers are positioned within a workplace that produces AI systems, how their position informs their claims, and the place of collective action in staking their claims. This chapter argues that workers occupy a unique role in identifying and mitigating harms caused by AI systems.

## 1. Introduction

AI is a social technology. It is designed by teams of developers who are employed by corporations, governments, academic institutions, or non-profit organizations. Using massive amounts of existing data, the technology learns from what it is fed (Valiant, 1984). Once implemented, it reflects—and automates—the biases that come both from its designers and the data it was trained on (Denton, et al., 2020). Without careful consideration of how this technology is designed and integrated into the systems we use, it can reproduce the biases and inequities that occur in everyday life (Friedman and Nissenbaum, 1996; Buolamwini and Gebru, 2018; Noble, 2018; Benjamin, 2019). As AI systems get deployed at an ever faster rate, affecting millions of people around the world, the scale at which such harms are propagated can be enormous (Crawford, 2021). Communities around the world– consumers, researchers, activists, and workers– have taken steps to understand and counteract the repercussions of these technologies on their lives. The field of AI governance has grown in tandem with AI technology with the goal of designing systems and protocols to minimize the potential harm of AI technologies (Dafoe, 2018).

While the role of states, corporations, and international organizations in AI governance has been extensively theorized, the role of workers in AI governance has received comparatively little attention. This chapter looks at the role that workers play in defining and mitigating harm that comes from AI technologies. We focus on societal harms as a category distinct from system reliability failures (Lynch and Veland, 2018). Harms are the causally assessed "impacts" of technologies (Moss, et al. 2021). They are not a result of technical negligence or a lack of oversight, but rather of normative uncertainty around questions of safety and fairness in complex social systems (Altman, Wood, and Vayena, 2019; Raji et al., 2020; Dobbe, Gilbert, Mintz, 2020). Numerous AI ethics frameworks conceptualize workers as stakeholders with a professional and personal interest in upholding ethics and preventing harms (Jobin, Ienca, Vayena, 2019). Within workplaces, however, the ability of workers to act as stakeholders is dependent on a number of factors. Examining the past decade of AI-related worker activism allows us to understand how different types of workers are positioned in workplaces that produce or propagate AI systems, how their position informs their claims, and the place of collective action in staking these claims.

Over the past decade, organizations that produce AI systems have sought to address these harms by establishing internal divisions for AI governance (cf. Google, 2021; Microsoft, 2021). These divisions operate as an internal reporting channel through which workers can report harms (Rakova, Yang, Cramer, and Chowdhury, 2021). They also contribute to a growing professional field of AI ethics that generates frameworks for diagnosing harms and treatments for addressing them. In some cases, these internal structures function as expected, and reports of harm are acknowledged and handled within the workplace. However, in other notable cases, workers and management disagree on how harms are identified or treated and workers turn to collective action to stake their claim on questions of AI governance (Belfield, 2020; Tarnoff, 2020).

In this chapter, we construct a model of harm reporting practices to explain how and why workers turn to collective action as a response. We draw on archival data of 25 collective actions publicly reported in the U.S. AI industry from 2010 to 2020 to inform our model (Tan and Nedzhvetskaya, 2021). We divide the harm reporting process into three steps (identification, governance decision, and response) with three stakeholders in the process (management, AI workers, and AI ethicists). Notably, our definition of AI workers includes both “designers”, workers who design, construct, or apply an AI technology or its surrounding infrastructure, and “trainers”, workers who feed data into an existing AI technology, supplement AI systems with human labor, or whose work is controlled by AI systems. Our archival data also shows that workers draw upon three types of claims to call for jurisdiction over questions of AI governance: subjection, control over the product of one’s labor, and proximate knowledge of systems. Workers turn to collective action to challenge higher-level decisions in the identification, governance decision, and response processes. Through collective action, workers lay claim to their own perspective on AI governance in the workplace. We examine these claims in later sections of the paper and evaluate how they may differ across categories of AI workers.

## 2. Who Is an AI Worker? Who Is an AI Ethicist?

To explore the role of workers in AI governance, it is necessary to begin with a definition of an AI worker. An AI worker must be employed or contracted by an institution that produces or uses AI systems. This could include public or private companies, academic institutions, and non-profit organizations. AI workers fall into two categories, “designers” and “trainers”.[1] The first category, “designers”, are workers who design, construct, or apply an AI technology or its surrounding infrastructure. This includes AI researchers, as well as those involved in developing and deploying AI systems, such as computer scientists and programmers. The second category, “trainers”, are workers who feed data into an existing AI technology, supplement AI systems with human labor, or whose work is controlled by AI systems. This includes workers who are reimbursed for the data they provide to train AI technologies– for example, rideshare drivers or app-based shoppers– and “artificial artificial intelligence” workers who supplement AI systems that would otherwise fail to perform “intelligently” (Shestakofsky and Kelkar, 2020).[2]

[1] We derive these terms from the processes laid out in Dobbe, Gilbert, and Mintz, 2020, though we are aware others have made this distinction or very similar ones.

The labor of “trainers" is as important to the design and function of algorithms as the code written by “designers”, and this has led to critiques of this binary (Amrute, 2019). We agree with this critique but choose to use the categories of “designer” and “trainer” because their entanglement with power dynamics explains some of the patterns we see in how these two categories of workers are treated across workplaces. This, in turn, shows why certain workers are more likely to make certain claims to AI governance. For example, AI trainers are more likely to be subject to harm from the algorithm itself and to use this claim when organizing a collective action. These two categories are entangled with power dynamics within institutions and within society more generally. Perceived value on the labor market can translate to differential treatment within the workplace (Rosenblat, 2019; Belfield, 2020). In many workplaces, divisions between these two categories of workers reflect existing inequalities in society along the lines of socio-economic class, race, gender, and nationality (Benjamin, 2019; Gray and Suri, 2019; Jack and Avle, 2021).

Another group of stakeholders that we must consider are AI ethicists: those involved in understanding and critiquing AI systems with the goal of reducing AI harms. While an AI ethicist may share some of the skills that AI workers have, such as implementing or deploying AI systems, they focus their time not on implementation, but on researching these systems for their potential to cause harm. Their expertise is not solely the result of professionalization, e.g.

[2] An important distinction must be drawn between AI workers of this category and AI consumers. Consumers use an AI-powered product (e.g. search engine) to fulfill some primary function (e.g. query for information) and provide their behavioral data to companies in conjunction with usage. They are reimbursed solely through access to a product. The “gift economy” business model, whereby consumers receive “free” access to products in exchange for their data, drives many of the world’s most profitable companies (Hoofnagle and Whittington, 2014; Fourcade and Kluttz, 2020). Despite the tremendous value they create, we do not consider these consumers in the category of AI workers.

specialized training or professional designations. Rather, expertise is a complicated arrangement of abstract and experiential knowledge, social ties, and institutional legitimacy (Eyal, 2013). AI ethicists spend time in different spaces (e.g. ethics conferences), establish different social ties (e.g. with fellow AI ethicists across companies, research centers, and universities), accumulate different experiences, and bear different institutional pressures than AI workers. Thus AI ethicists form their own "epistemic community" with a shared set of experiences distinct from AI workers (Haas, 1992). Nevertheless, internal AI ethicists are employees in the same respect that AI workers are employees and face similar pressures to prove their value to their employer and comply with the norms and goals of their institution.

Institutions have many reasons for hiring experts such as AI ethicists. Expertise is a claim to jurisdiction over some field of abstract knowledge associated with a set of concrete tasks (Eyal, 2013). Institutions seek to ensure their own survival and reflect this in their actions (Berger and Luckmann, 1966; Meyer and Rowan, 1977). Experts can bring legitimacy to an institution, increasing their chances of survival by strengthening support for their actions both internally and externally. Notably, experts are valued not for their ability to apply objective facts in uncontested situations but rather for their ability to use complex inference to solve problems in contested domains (Abbott, 1988). When institutions hire experts for their expertise and retain oversight of their work, they maintain jurisdiction for the institution itself to the extent that this jurisdiction does not encroach on external professional or legal codes. For example, an institution may hire a lawyer for their legal expertise. The lawyer will work on behalf of the institution but within the realm of legal code. By hiring internal ethicists, institutions maintain control over how the abstract knowledge of expertise can be applied to critique the institution's operations.

This is especially the case in the field of AI ethics. Because of the nature of the field, AI ethicists' expertise extends not only to empirical knowledge but also to moral questions, which can make them a crucial resource for institutions seeking legitimacy for their AI products and services (Bostrom and Yudkowsky, 2011; Greene, Hoffman, and Stark, 2019). As with other forms of business ethics, AI ethics provides a moral background for the field of AI, a set of causally informed moral concepts that allow us to establish norms that inform our behaviors (Abend, 2014). Institutions hire AI ethicists for their jurisdiction over what is contested, uncertain, and subjective when it comes to AI harms. In the following section, we demonstrate how ethicists are integrated in AI governance in the workplace and the ways in which they must simultaneously alternate between their roles as experts and workers. We begin by introducing the data that informed our model before explaining the harm reporting process in AI workplaces.

## 3. Harm Reporting in AI Workplaces

Our model of harm reporting practices is limited to workplaces that produce AI systems, or what we will call an "AI workplace" from here on. To construct this model, we draw upon archival data on collective actions in the technology industry from 2010 to 2020. Of the over 300 actions catalogued, 25 directly mention or involve AI. Table 1 provides a chronological list of these collective actions. From here on, we refer to these as *AI collective actions*. In our harm reporting model, we divide the institution's process for evaluating and managing harm into three parts–identification, governance decision, and response–and identify the roles played by management, AI workers, and AI ethicists (Table 2). Notably, this framework applies to both "designers" and "trainers".

To construct our archive on collective actions in the technology industry, we gathered data using NexisUni news archives. We searched for articles where collective action terms[3] occurred within 25 words of employment terms[4] for the computing & information technology industry. To maintain the archive, we draw from a variety of sources including but not limited to a network of active organizers in the tech industry, reporters on the tech-labor beat, and Google Alerts. To qualify for our archive, events must be "collective"[5] and present "evidence of action"[6] by currently or recently employed "tech workers".[7] For this analysis, we only include actions that directly mention or involve AI, protesting the use of an AI technology, the application of AI to govern or control their own labor, or the stakes of AI governance in the workplace.

Our archive of events is limited only to actions that have been reported by a news publication and is heavily skewed through actions reported in the English language press since we conducted the search only in English. In addition, news publications can be biased in their reporting of social movement actions based on their geographic locale and political orientation (Davenport, 2009). We believe this to be the case for collective actions in the technology industry as well. 72% of actions covered in our archive through 2020 occurred in the United

---

[3] Collective action terms include: protest*, petition*, strike*, open letter*, walk out*, union*, boycott*, letter*, lawsuit*, discuss*, negotiat*.

[4] Employment terms include: employee*, worker*, contract*, labor*.

[5] To be considered "collective", events must involve a minimum of two employees who recognize themselves as a group united by a shared cause and/or issue. The cause and/or issue should be relevant to a broader public, defined as a community which is not directly related to the company through employment or financial ties. The broader public does not include company shareholders, businesses with shared interests, or owners of company property, which include consumers of company products. An action that would typically not be deemed relevant to a broader public on its own merit may qualify if the group presents an argument that it is occurring in response to a cause and/or issue that is relevant to a broader public (e.g. an employee who is believed to have been fired as retaliation for a recent protest).

[6] To present "evidence of action", events must involve an attempt to present the cause and/or issue outside of the immediate group. Actions may be either internal (available or visible only to other employees) or external (available or visible to the broader public). Lawsuits may be included if they are granted class action status or if they incite additional collective action. Actions should not be initiated by company management.

[7] "Tech workers" are defined as those current or recently employed (within the last year) workers in the technology industry. The technology industry includes but is not limited to information technology, Internet, hardware, software. It does not include adjacent industries, e.g. digital media or the video game industry. It does include online retailers and social media companies. Academics whose research concerns technology and students or interns who are preparing to enter the tech industry can be considered tech workers.

States, Canada, or online and 84% were in the United States, Canada, Europe or online. 58% of actions took place in one of seven American corporations: Amazon, Apple, Facebook, Google, Lyft, Microsoft, and Uber. We have attempted to counteract these biases by opening the archive to crowdsourced contributions, which are more likely to come from outside of the United States and to include smaller companies and institutions. To date, approximately 5% of events have been contributed to the archive through crowdsourcing. In the following sections, we draw upon cases from Table 1 to illustrate how harm reporting practices have proceeded in AI workplaces.

<< Table 1 here>>

## 1. Identification

During *identification*, harms are identified by AI workers as part of their routine labor or by AI ethicists as part of an audit or research study.[8,9] The identification process can take place immediately upon encountering the harm or gradually over the process of becoming acquainted with the systems at work. Identification may preempt potential harms, as in the case of Microsoft employees that called on the company to abstain from bidding on the now-cancelled Joint Enterprise Defense Infrastructure (JEDI) contract from the U.S. Department of Defense (Employees of Microsoft, 2018). The JEDI contract is designed to accelerate the military's adoption of AI technology (Williams, 2020) and increase the military's lethality (Employees of Microsoft, 2018).

[8] Sometimes, harms are identified by the media and then acted upon by workers. The case of the Google Walkout is a good example; a New York Times report about the $90-million payout to Andy Rubin catalyzed the Google Walkout (Wakabayashi and Benner, 2018)

[9] Harms can be identified by management in theory, though we have not documented such an event in our archive. Research suggests that management is sensitive to reputational risks from AI harms and can be compelled to act to prevent reputational damage (Rakova, Yang, Cramer, and Chowdhury, 2021).

In other instances, harms have been identified years after their initial repercussions were felt. In 2020, British Uber drivers filed a lawsuit in Dutch courts alleging that the company's firing algorithm violated GDPR Article 22. More than one thousand individual cases of firings were presented dating two years back (Russon, 2020). While individual workers identified the harms much earlier, when they presented these harms to the company, the company failed to recognize them as such. By gathering cases together in a class action lawsuit, i.e. taking a collective action, workers were able to contest this decision. Among the two categories of AI workers described, trainers are more likely to find themselves the subject of harm. The nature of the work is itself often governed by algorithms and workers are often the first to identify harms as a result (Rosenblat and Stark, 2016).

## 2. Governance Decision

Once a harm is identified, the question of how to manage that harm, and who gets to decide, comes next. This leads us to the second step of the model, the *governance decision*. The *governance decision* is a way for the institution to decide how to manage an identified harm. Typically, decisions made in this section are controlled by management. Here, management must evaluate: is there an existing governance framework that determines how this harm should be dealt with? During the governance decision, management may consult with AI ethicists to see if there is an existing framework to diagnose the identified harm.

Conflicts can arise during the governance decision when recognizing a harm threatens the survival of an institution or reduces its legitimacy. Corporations, for instance, must demonstrate their present or future potential for profit in order to receive investment from their shareholders (Useem, 1993; Fligstein, 2001).When reported harms threaten the success of profitable products, managers within a corporation must decide between their obligation to generate profit for shareholders or their obligation to adhere to the ethical standards of the field. The two may coincide if management believes that a harm poses enough of a reputational risk to the firm that it threatens future profitability (Rakova, Yang, Cramer, and Chowdhury, 2021).

Among the 25 collective actions documented, ten protested specific corporate contracts. Conflicts between corporate pressures to generate profit and workers' personal and professional ethical standards are a common source of collective action in the AI workplace. In 2020, Dr. Timnit Gebru, a co-lead on Google's Ethical AI Team, was removed from the company after submitting an academic paper to management for review. The paper included critiques of a category of language processing models that are core to the company's highly profitable search engine business (Hao, 2020). The conditions of Dr. Gebru's removal from the company led individuals close to the case to allege that the action was a case of censorship and incited a number of collective actions from Google workers and the broader AI research community (Google Walkout for Real Change, 2020; Canon, 2020; Allyn, 2020).

Notably, concerns over lack of legitimacy triggered by worker collective actions can lead management to formalize ethics review processes, laying out objectives and bolstering AI ethicists on staff. In June 2018, following several months of internal activism against the company's involvement with Project Maven, an AI program contracted by the U.S. Department of Defense, Google CEO Sundar Pichai released a blog post announcing the company's new AI principles (Shane, Metz, and Wakabayashi, 2018; Pichai, 2018). In the post, Pichai stated that the company would not "design or deploy AI...that cause or are likely to cause overall harm. Where there is a material risk of harm, we will proceed only where we believe that the benefits substantially outweigh the risks, and will incorporate appropriate safety constraints."

Formalizing review processes and articulating ethical principles can improve accountability, but does not address the issue of oversight and objectivity in internal governance without accompanying organizational change (Posada, 2020; Amrute, 2021; Rakova, Yang, Cramer, and Chowdhury, 2021).

3. Response

The *response* can take a few forms. When a framework exists to diagnose the harm and there is high consensus among all stakeholders, the harm can be treated through the prescribed method: management consults with AI ethicists, and AI ethicists are able to diagnose and treat the harm. Treatment of the harm can also involve AI workers. "Trainers" may have to re-label data or collect data in a different way in order to mitigate the identified harm. "Designers" may have to alter the nature of the product itself. Treatment itself can generate conflict, however, if stakeholders disagree about its application or feel it is insufficient to address the scale of the harm. Lack of consensus can occur when management disagrees with AI ethicists' diagnosis of the harm or dismisses or reclassifies the harm. In cases where an AI ethicist is unable to exercise their expertise, they must participate as a worker to stake their claim.

If there is low consensus among stakeholders about the existing framework or its prescribed treatment, AI workers may choose to engage in individual action (e.g. whistleblowing) or collective action as part of the response. Collective action is defined by social movement theorists as "emergent and minimally coordinated action by two or more people that is motivated by a desire to change some aspect of social life or to resist changes proposed by others" (McAdam, 2015). While this definition is broad enough to encompass all stakeholders involved in our model– management, AI workers, and AI ethicists— collective action is seldom taken by management because managers have enough power within the institution to act individually. For AI workers, what is crucial, is that the power to incite or resist change is partly the result of acting as a collective, i.e. claiming to represent a viewpoint greater than the individual.

AI workers are more likely to engage in collective action when other means of influence are not available to them or have proven insufficient. Through collective action they can contest an identification process or a governance decision and gather support, either internally within the company or externally with a broader public, for a different outcome. Critically, such actions do not come without consequences, and there have been numerous accounts of retaliation towards AI workers that have spoken out against their employers.[10] Fundamentally, these actions are political in that they challenge existing power structures in the workplace. Importantly, AI ethicists, as well as AI workers, may engage in collective action. In the following section, we examine some of the ways in which workers make these claims.

<< Table 2 here>>

**4. Worker Claims**

AI workers turn to collective action to challenge higher-level decisions in the identification, governance decision, and response processes in the AI workplace. Through collective action, AI workers stake their own claims on questions of governance. In the AI collective actions we have identified, there are three primary ways in which AI workers construct their claims to governance: subjection, control over the product of one's labor, and proximate knowledge of a system (Table 3).

---

[10] Former Google employees, Meredith Whittaker and Claire Stapleton, were retaliated against by Google after publicly voicing criticism of their employer (Tiku, 2019). Amazon fired Emily Cunningham and Maren Costa after their public criticism of Amazon's handling of Covid-19 and poor environmental practices, which included criticism of Amazon's sale of AI capabilities to oil companies (Greene, 2020). Though the actions taken are collective, the risks and consequences of organizing such actions are often faced by individuals (Amrute, 2021).

First, AI workers themselves may claim to be subject to the harms they identify. In the case of delivery workers whose work is controlled by AI-powered algorithms, workers themselves experience the harms of exploitative systems that are designed to extract the greatest possible amount of labor out of employees, for example, incentivizing them towards overwork (Rosenblat and Stark, 2016; Lei, 2021). Delivery workers in China offer a good example of AI workers identifying and resisting AI-powered algorithmic management. Ele.me and Meituan, the two largest food delivery platforms in China, each have their own AI systems that, according to their creators, reduce the time for deliveries and give them an advantage over competitors.[11] However, from the perspective of the delivery drivers, these systems simply pass the burden of making more efficient deliveries onto drivers (Renwu Magazine, 2020). Drivers have demonstrated how the apps instruct them to take more dangerous routes in order to make a delivery on time, for example, advising them to ride against traffic on a one-way street. In response, delivery workers have participated in dozens of strikes, directly or indirectly protesting the harms they have faced as a result of the algorithm. According to the China Labor Bulletin, there were 45 recorded strikes in 2019 by food delivery drivers in China.[12]

Notably, AI workers may also identify with communities that are disproportionately impacted by the harms they identify, for example, vulnerable immigrant populations or racial targets of incarceration who are more likely to be subject to surveillance technologies (Selbst, 2017; Benjamin, 2019). Many of the leaders of collective actions in the technology industry have been women, people of color, or have identified as LGBTQ (Tarnoff, 2020). This is particularly salient for AI designers who themselves may not be the subject of harm but identify with particular communities or social movements. During the Black Lives Matter movement in 2020, many tech workers who identified as activists in the broader movement for racial justice raised the issue of police brutality and surveillance in their workplaces. Microsoft employees—who drew on their own experiences and participation in the movement—wrote and collectively signed an email to Microsoft's leaders calling for the company to cancel its contract with the Seattle Police Department (SPD). "Every one of us in the CC line are either first hand witnesses or direct victims to the inhumane responses of SPD to peaceful protesting" the email states (Gershgorn, 2020). While the explicit demand to cancel the contract with the SPD was ignored, Microsoft banned the use of their facial recognition software the following month.

[11] Ele.me's AI system is named "Ark." According to its engineers, Ark is able to adjust and optimize deliveries for lunch hours vs regular hours and high rises vs small neighborhoods. (Si, 2018) Meituan's equivalent system is called "Super Brain." These systems are allegedly able to help delivery drivers save time by providing them with smarter routes and facilitating a more efficient distribution of orders across the system.
[12] China Labor Bulletin recorded 45 strikes by delivery drivers in 2019 (China Labor Bulletin Strike Map. 2020)

Second, AI workers may contest how the products of their own labor are being applied. They may identify sufficiently with the product of their labor that they feel responsible for determining that it goes towards an end they view as beneficial or optimal (Caves, 2000; Ranganathan, 2018). In the aforementioned 2018 open letter by Microsoft employees asking management to drop a bid for the Department of Defense's JEDI contract, employees drew comparisons across ethics-driven activism across technology firms. "Like those who took action at Google, Salesforce, and Amazon, we ask all employees of tech companies to ask how your work will be used, where it will be applied, and act according to your principles," the letter stated (Employees of Microsoft, 2018). Such themes reverberate in other open letters, particularly those involving social and ethical issues, such as climate change (Amazon Employees for Climate Justice, 2019; Tech Won't Drill It, 2020; Microsoft Employees, 2019).[13]

[13] "We, members of the AI and physical sciences communities, are distressed to find that the intellectual output of our research, which has been performed with the goal of benefiting humanity, will be applied to further deepen the climate crisis. These applications are not only misaligned with our values and moral responsibilities, but they are also in direct conflict with the

Notably, both AI designers and AI trainers may claim control over the product of one's labor. Benchmark machine learning datasets–and the trainers who have coded them– play a foundational role in determining how AI systems understand abstract social concepts. The biases and structural limitations of these datasets create epistemic commitments, i.e. "genealogies of data", that proliferate as adoption of these benchmark datasets grows (Denton, et al. 2020). For AI ethicists, the product of their labor is the identification and treatment of the harm itself and they may launch this claim if they feel a harm has not been sufficiently acknowledged or addressed.

Third, AI workers may have a privileged insight into the workings and consequences of the products they create by virtue of the time they spend constructing them, i.e. they have proximate and privileged knowledge of the systems (Perrow, 1984). In 1986, a group called Computer Professionals for Social Responsibility (CPSR), which included fourteen AT&T employees and thirty external computer scientists from industry and academia, wrote a letter to U.S. President Ronald Reagan alleging that Star Wars, a software system intended to defend the U.S. against ballistic missiles, was too error-prone to responsibly run (Boffey, 1986). Much of the group's argument drew upon their combined expertise in software development. They cited the inevitability of "bugs" in software code, the difficulty of fully testing software built at such a large scale, and the system's reliance on tightly coupled feedback loops. [14]

While AI technology was not implemented as part of the "Star Wars" system in 1986, the focus on system complexity and safety has been a consistent theme in studies of AI safety and AI accidents (Amodei, et al., 2016; Maas, 2018). AI accidents are defined as "harmful behavior[s]"

---

existential interests of planetary life." (Tech Won't Drill It, 2020) "We have to take responsibility for the impact that our business has on the planet and on people." (Amazon Employees for Climate Justice, 2019).

[14] Since its inception, researchers have noted how software is especially error-prone compared to more physical technologies (Parnas, et al., 1990). Industry estimates assume approximately 15 to 50 errors per every thousand lines of non-AI computer code (McConnell, 2004).

that result from a misspecification, a misunderstanding, or an oversight of AI processes (Amodei, et al., 2016). AI accidents can be considered part of a larger genre of “system accidents”, also called “normal accidents”, which are the result of mistakes that rapidly escalate as a result of tight coupling and feedback loops in systems (Perrow, 1984). Machine learning, neural networks, and many other general purpose AI technologies rely on exactly this structure of tightly coupled feedback loops and are highly prone to system accidents as a result (Maas, 2018; Fourcade and Johns, 2020). System accidents are difficult to understand and challenging to prevent. In systems where no single individual has full knowledge of the system or how its component parts interact, full oversight is impossible, even by highly trained experts (Elish, 2019).

One means of preventing these accidents is to allow sufficient decentralization so that AI workers can factor in localized variables that may impact operations and respond accordingly (Vaughan, 1996). AI workers are often the only individuals with sufficient proximity to these systems to act proactively in calling out potential accidents or harms. Institutions designing AI systems can purposefully make elements of these systems opaque as a means of “self-protection”, guarding against industry competition or public scrutiny (Pasquale, 2015; Burrell, 2016). Scholars have demonstrated how dissent channels that draw upon democratic consensus-building can be integrated within AI design, training, and deployment processes (Dobbe, Gilbert, and Mintz, 2019). Such mechanisms, however, cannot function properly under the jurisdiction of a single institution, which exist to satisfy their own institutional goals as opposed to greater societal objectives. External arbitrators need to be involved to specify terms and enforce negotiations in a way that can maintain trust and cooperation (Hadfield-Menell and Hadfield, 2018). Collective action has been an important mechanism for AI workers to voice their dissent in the workplace when external arbitrators are not available.

<< Table 3 here>>

## 5. Conclusion

In this chapter, we describe how AI harms are handled within institutions and demonstrate how the positionality of management, ethicists, and workers affect the treatment of the identified harm. Each of these roles comes with its own set of responsibilities for designing and managing AI systems. These institutional roles interact with an individual's existing ethical beliefs, lived experiences, and interests to allow them to come to their own judgment of what constitutes a harm and how best to mitigate this harm.

There are three steps involved in the harm reporting process: identification, the governance decision, and the response. Workers turn to collective action as a response to challenge higher-level decisions when they lack the individual power within their own role. The claims workers make are based on their relationship to the system they are critiquing. Workers may claim they are themselves subject to harm as a result of an AI system. If they played a role in designing or training the system itself, workers may claim control over the product of their own labor or proximate knowledge of the system. Through collective action, AI workers stake a claim for themselves in AI governance in the workplace and assert their perspective as individuals whose labor is entangled with these systems.

Thus far, research in AI governance and labor has focused on the identification stage of the harm reporting process, particularly on providing frameworks by which practitioners can identify and evaluate harms. There has been an open acknowledgment of the ways in which identification is an inherently political process, one that requires questioning existing power structures and inequalities within society. Outlining the two additional steps that follow– the governance decision and the response– allows us to trace the full political process of harm reporting in the AI workplace and identify where power lies within this process. There is a path dependency in this process that is important to acknowledge. How harms are identified can determine how harms are governed and treated. As research progresses to these second and third steps, it will be necessary to capture and describe these links between identification, governance decision, and response. The model presented here provides a generalizable framework which can enable this research.

*Acknowledgments:* The authors would like to thank April Mariko Salazar for excellent research assistance. This research was supported by funding from the Jain Family Institute and the Center for Technology, Society, and Policy at the University of California, Berkeley. Earlier versions of this paper benefited from feedback provided by Baobao Zhang, Francis Tseng, Rumman Chowdhury, members of the Jain Family Institute Digital Ethics Reading and Works-in-Progress Group, and participants in the 2021 Genealogies of Data Workshop.

## Appendix

Table 1. Public Collective Actions in AI, 2010 to 2010

| Date | Institution | Collective Action | # | Claims | Worker Type | Description |
|---|---|---|---|---|---|---|
| Feb. 2013 | Amazon | New Platform | NA | Subjection; Proximate Knowledge | Trainers; Designers | Mechanical Turk workers have adopted a platform named Turkopticon, which allows them to review tasks and requesters on the platform. |
| Sept. 2014 | Amazon | Open Letter; Standard Setting | 262 | Proximate Knowledge; Product of One's Labor; Subjection | Trainers; Designers | Mechanical Turk workers and academics, organized through Dynamo, have collaborated and signed a set of ethical guidelines for scholars requesting labor through the platform. |
| Dec. 2014 | Amazon | Open Letter | 550 | Subjection | Trainers | Mechanical Turk workers participate in an email writing campaign to Amazon founder Jeff Bezos to protest low pay and worker representation. |
| Aug. 2017 | Amazon | New Platform | 550 | Proximate Knowledge; Product of One's Labor; Subjection | Trainers; Designers | Mechanical Turk workers and organizers collaborated with researchers at Stanford's Crowd Research Collective to create Daemo, a crowd-sourced platform that provides a higher-paying alternative to Amazon's Mechanical Turk. |
| April 2018 | Academic Institutions | Open Letter | 50 | Product of One's Labor; Proximate Knowledge | Designers; Ethicists | A group of academics working on artificial technologies call for a boycott of the Korea Advanced Institute of Science and Technology, a South Korean university believed to be developing AI weapons through a collaboration with a defense contractor. |
| May 2018 | Google | Open Letter | 5000 | Product of One's Labor | Mostly Designers | Google employees have led a campaign demanding that their company terminate its contract with the Pentagon for Project Maven, a program that uses machine learning to improve targeting for drone strikes. |
| June 2018 | Google | Internal Protest | 9 | Product of One's Labor | Designers | A group of influential software engineers in Google's cloud division, referred to as the 'Group of Nine', surprised their superiors by refusing to work on a cutting-edge security feature. Known as "air gap," the technology would have helped Google win sensitive military contracts. |
| July 2018 | Microsoft | Open Letter | 500 | Product of One's Labor | Mostly Designers | Two Microsoft employees presented their CEO with an online petition including more than 300,000 signatures (including 500 employees) calling for the firm to cancel their contract with the US Immigration and Customs Enforcement Agency. |
| Oct. 2018 | Microsoft | Open Letter | NA | Product of One's Labor | Designers | An open letter signed by 'employees of Microsoft' asks the cloud giant to abstain from bidding on the military's massive JEDI cloud computing contract for ethical reasons involving the application of artificial intelligence. |
| Feb. 2019 | Microsoft | Open Letter | 290 | Product of One's Labor | Designers | Microsoft employees circulated a letter among the companies' over 130,000-person staff demanding that executives cancel a $479 million contract with the US Army, IVAS, that will provide weapons technology to the U.S. Military to 'increase lethality' to army soldiers. |
| March 2019 | Amazon | Open Letter | 45 | Product of One's Labor | Designers; Ethicists | Researchers are calling on Amazon to stop selling facial recognition software to law enforcement. |
| April 2019 | Google | Open Letter | 2556 | Product of One's Labor | Mostly Designers | Google employees, along with academic, civil society, and industry supporters, have called for the removal of a rightwing thinktank leader from the companies's new artificial intelligence council, citing her anti-LGBT and anti-immigrant record. |
| May 2019 | Facebook | Internal Protest | 12 | Subjection | Trainers | Facebook moderators in the United States have been spearheading a quiet campaign inside the social media giant to air their grievances about unsatisfactory working conditions and their status as second-class citizens. |
| Nov. 2019 | Instacart | Open Letter; Protest | 212 | Subjection; Proximate Knowledge | Trainers | Instacart workers circulated an open letter on Medium prior to a walkout. Employees are protesting the company's history of systematically devaluing labor, using algorithms to reduce pay to employees and changing tipping structures to return profit to the company. |
| Nov. 2019 | Google | Open Letter | 1137 | Product of One's Labor | Designers; Ethicists | Google employees have signed a letter calling for their employer to take action on the issues of climate change and immigration. The letter called on Google to commit to zero carbon emissions by 2030, end all contracts with fossil fuel companies and climate change denying or delaying 'think tanks, lobbyists, and politicians', and end all collaboration with organizations or individuals 'enabling the incarceration, surveillance, displacement, or oppression of refugees or frontline communities'. |
| 2019 | Meituan, Ele.me, Fengniao | Strikes | NA | Subjection | Trainers | In 2019, food delivery drivers have participated in over 45 strikes across China over issues including demanding higher pay and protesting change in distance calculation methods |
| Feb. 2020 | Academic Institutions | Open Letter | 141 | Product of One's Labor | Designers; Ethicists | AI researchers and engineers from around the world signed an online open letter calling for the tech industry to refrain from using AI technologies to 'exacerbate the climate crisis'. |
| June 2020 | Google | Open Letter | 1650 | Product of One's Labor | Designers | Google employees have signed an open letter to CEO Sundar Pichai demanding the company stop selling its technology to police forces across the US. |
| June 2020 | Academic Institutions | Open Letter | 600 | Proximate Knowledge | Designers; Ethicists | Hundreds of expert researchers and practitioners across a variety of technical, scientific, and humanistic fields (including statistics, machine learning and artificial intelligence, law, sociology, history, communication studies and anthropology) sign a letter calling for a forthcoming publication entitled “A Deep Neural Network Model to Predict Criminality Using Image Processing”, to be rescinded from publication. |
| June 2020 | Microsoft | Open Letter | 250 | Subjection | Designers | Microsoft employees wrote and collectively signed an email to Microsoft’s leaders calling for the company to cancel its contract with the Seattle Police Department. In the following month, Microsoft banned the use of their facial recognition software. |
| Oct. 2020 | Facebook | Open Letter | 290 | Subjection | Designers; Trainers | Facebook employees and contractors signed a petition demanding a 50% wage increase as hazard pay for site moderators who have been forced to return to the office by October 12th, 2020. A Facebook spokesperson said the reason for the return was that "some of the most sensitive content can't be reviewed from home". |
| Oct. 2020 | Uber | Legal Action | 1000 | Subjection | Trainers | British Uber drivers are launching a lawsuit in the Netherlands to protest the company's use of algorithms to fire workers. |
| Nov. 2020 | Facebook | Open Letter | 200 | Subjection | Trainers; Designers | Facebook workers signed a letter demanding better treatment for content moderators after being required to return to the office in the midst of the pandemic. |
| Dec. 2020 | Google; Academic Institutions | Open Letter | 2695 | Product of One's Labor; Proximate Knowledge | Designers; Ethicists | After Dr. Timnit Gebru was fired from Google, Googlers wrote an open letter demanding accountability from Google. |
| Dec. 2020 | Google | Open Letter | NA | Product of One's Labor; Proximate Knowledge | Ethicists | Two weeks after Dr Timnit Gebru was fired from her role in Google, Google's Ethical AI employees sent CEO Sundar Pichai a list of demands, including organizational changes and a request to reinstate dismissed researcher Timnit Gebru at a higher level. |

Table 2. Harm Reporting Process Model

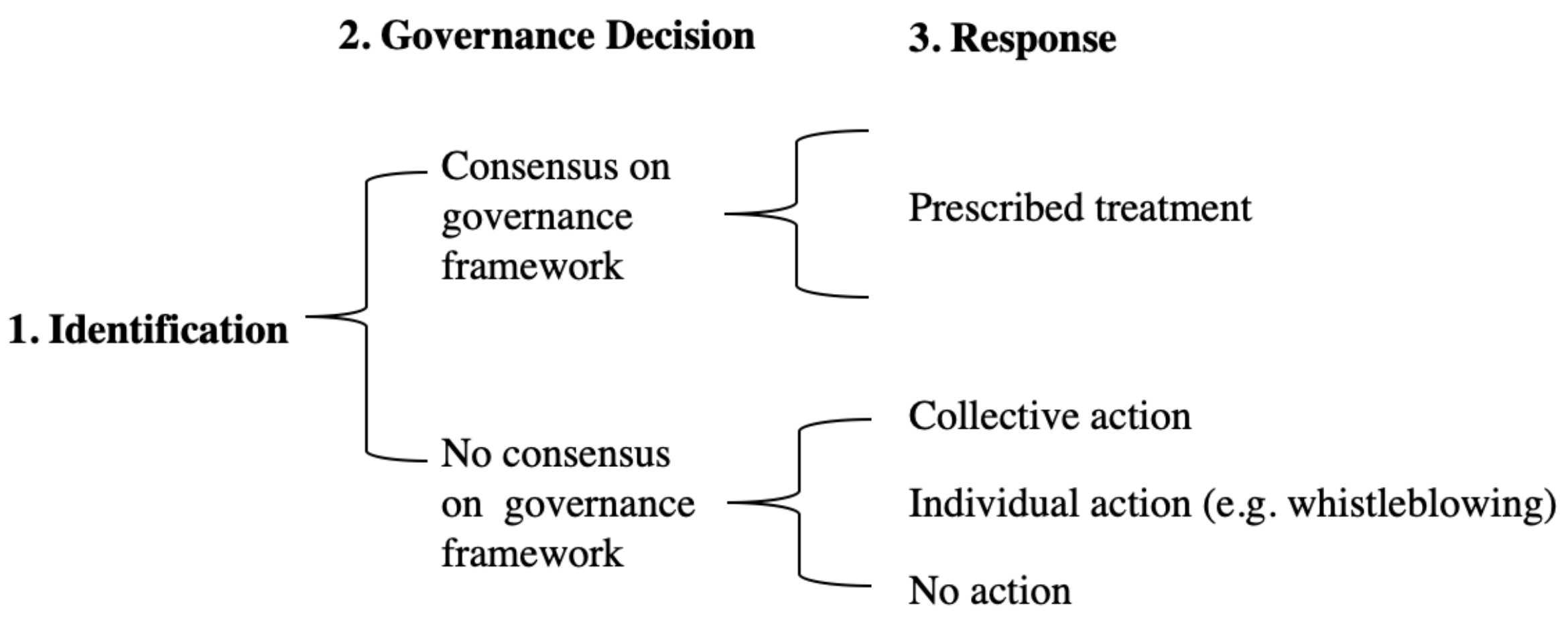

Table 3: Types of Worker Claims

| Claim | Description | AI Worker Types | Example |
|---|---|---|---|
| Subjection | Workers themselves are subject to the harms they identify within AI systems | Trainers more vulnerable to subjection | British Uber drivers filed a lawsuit in the Netherlands to protest the company's use of algorithms to fire workers (Oct. 2020). |
| Product of One's Labor | Workers contest how the products of their own labor are being applied | Both types of workers may claim | An open letter signed by 'employees of Microsoft' asks the company to abstain from bidding on the military's massive JEDI cloud computing contract for ethical reasons involving the application of AI to warfare (March 2019). |
| Proximate Knowledge | Workers claim a privileged insight into the workings and consequences of the AI systems they create by virtue of their proximity | Both types of workers may claim | Mechanical Turk workers and academics, organized through Dynamo, have collaborated and signed a set of ethical guidelines for scholars requesting labor through the platform (Sept. 2014). |

**Bibliography**

Abbott, Andrew. 1988. *A System of Professions*. Chicago: University of Chicago Press.

Abend, Gabriel. 2014. *The Moral Background: An Inquiry into the History of Business Ethics*. Princeton: Princeton University Press.

Altman, Micah, Wood, Alexandra, and Effy Vayena. 2018. "A Harm-Reduction Framework for Algorithmic Fairness." *IEEE Security Privacy* 16(3):34-45.

Allyn, Bobby. 2020. "Google AI Team Demands Ousted Black Researcher Be Rehired And Promoted." *National Public Radio.* https://www.npr.org/2020/12/17/947413170/google-ai-team-demands-ousted-black-researcher-be-rehired-and-promoted

Amazon Employees for Climate Justice. 2019. "Amazon employees are joining the Global Climate Walkout, 9/20". *Medium.* https://amazonemployees4climatejustice.medium.com/amazon-employees-are-joining-the-global-climate-walkout-9-20-9bfa4cbb1ce3

Amodei, Dario and Olah, Chris and Steinhardt, Jacob and Christiano, Paul and Schulman, John and Mané, Dan. 2016. "Concrete problems in AI safety." arXiv preprint arXiv:1606.06565.

Amrute, Sareeta. 2019. "Of Techno-Ethics and Techno-Affects." *Feminist Review*. 123:56-73.

Amrute, Sareeta. 2021. "A New AI Lexion: Dissent." *AI Now Institute*. https://medium.com/a-new-ai-lexicon/a-new-ai-lexicon-dissent-2b7861cad5ff

Angwin, Julia, Jeff Larson, Surya Mattu, and Lauren Kirchner. 2016. "Machine bias." *ProPublica.* 139-159.

Barabas, Chelsea, Colin Doyle, JB Rubinovitz, and Karthik Dinakar. 2020. *Studying up: Reorienting the Study of Algorithmic Fairness around Issues of Power Proceedings of the 2020 Conference on Fairness, Accountability, and Transparency.* Presentation. FAT '20. New York, NY, USA Barcelona, Spain: Association for Computing Machinery.

Belfield, Haydn. 2020. "Activism by the AI Community: Analysing Recent Achievements and Future Prospects." Presentation. AIES '20. New York, NY, USA: Association for Computing Machinery.
Benjamin, Ruha. 2019. *Race After Technology: Abolitionist Tools for the New Jim Code.* New York: Polity.

Berger, Peter L. and Thomas Luckmann. 1966. *The Social Construction of Reality: A Treatise in the Sociology of Knowledge*. London: Penguin Books.

Boffey, Philip M. 1986. "Software Seen as Obstacle in Developing 'Star Wars'". *The New York Times.* https://www.nytimes.com/1986/09/16/science/software-seen-as-obstacle-in-developing-star-wars.html

Bostrom, Nick. 2011. *Superintelligence: Paths, Dangers, Strategies.* Oxford: Oxford University Press.

Buolamwini, Joy and Timnit Gebru. 2018. "Gender Shades: Intersectional Accuracy Disparities in Commercial Gender Classification." *Proceedings of the 1st Conference on Fairness, Accountability and Transparency.* Presentation. Proceedings of Machine Learning Research. New York, NY, USA: PMLR.

Burrell, Jenna. 2016. “How the machine ‘thinks’: Understanding opacity in machine learning algorithms.” *Big Data & Society.* 3(1).

Canon, Gabrielle. 2020. “Google workers reject company's account of AI researcher's exit as anger grows.” *The Guardian.* https://www.theguardian.com/technology/2020/dec/07/timnit-gebru-google-firing-resignation-ai-research

Caves, Richard E. 2000. *Creative industries: Contracts between art and commerce*. Cambridge, MA: Harvard University Press.

China Labor Bulletin Strike Map. 2020. China Labor Bulletin. https://maps.clb.org.hk/?i18n_language=en_US&map=1&startDate=2019-01&endDate=2019-12&eventId=&keyword=&addressId=&parentAddressId=&address=&industry=10308&parentIndustry=&industryName=Food%20delivery

Coalition for Critical Technology. 2020. “Abolish the #TechToPrison Pipeline.” *Medium.* https://medium.com/@CoalitionForCriticalTechnology/abolish-the-techtoprisonpipeline-9b5b14366b16

Crawford, Kate. 2021. *Atlas of AI: Power, Politics, and the Planetary Costs of Artificial Intelligence.* New Haven: Yale University Press.

Dafoe, Allan. 2018. *AI governance: a research agenda*. Oxford: University of Oxford.

Davenport, Christian. 2009. *Media Bias, Perspective, and State Repression: The Black Panther Party*. Cambridge: Cambridge University Press.

Denton, Emily, Hanna, Alex, Amironesei, Razvan, Smart, Andrew, Nicole, Hilary and Morgan Klaus Scheuerman. 2020. “Bringing the People Back In: Contesting Benchmark Machine Learning Datasets.” *arXiv.* 2007.07399.

Dobbe, Roel I.J., Thomas Krendl Gilbert, and Yonatan Mintz. 2020. *Hard Choices in Artificial Intelligence: Addressing Normative Uncertainty through Sociotechnical Commitments Proceedings of the AAAI/ACM Conference on AI, Ethics, and Society.* Presentation. AIES ‘20. New York, NY, USA: Association for Computing Machinery.

Elias, Jennifer. 2021. “Another prominent Google scientist is leaving the company amid fallout from fired AI researcher.” *CNBC.* https://www.cnbc.com/2021/04/06/googles-samy-bengio-is-leaving-amid-fallout-from-ai-researcher-firing.html

Elish, Madeleine Clare. 2019. “Moral Crumple Zones: Cautionary Tales in Human-Robot Interaction.” *Engaging Science, Technology, and Society.* 5.

Employees of Microsoft. 2018. “An Open Letter to Microsoft: Don’t Bid on the US Military’s Project JEDI”. https://medium.com/s/story/an-open-letter-to-microsoft-dont-bid-on-the-us-military-s-project-jedi-7279338b7132

Eyal, Gil. 2013. “For a Sociology of Expertise: The Social Origins of the Autism Epidemic.” *American Journal of Sociology.* 118 (4):863-907.

Fligstein, Neil. 2001. *The Architecture of Markets: An Economic Sociology of Twenty-First Century Capitalist Societie*s. Princeton, NJ: Princeton University Press.

Fourcade, Marion and Fleur Johns. 2020. “Loops, ladders and links: the recursivity of social and machine learning.” *Theory and Society*. 49 (5):803-832.

Fourcade, Marion and Daniel N. Kluttz. 2020. “A Maussian bargain: Accumulation by gift in the digital economy.” *Big Data & Society*. 7 (1).

Friedman, Batya and Nissenbaum, Helen. 1996. “Bias in computer systems.” *ACM Transactions on Information Systems (TOIS).* 14 (3):330-347.

Gershgorn, Dave. 2020. “250 Microsoft Employees Call on CEO to Cancel Police Contracts and Support Defunding Seattle PD” *One Zero*. https://onezero.medium.com/250-microsoft-employees-call-on-ceo-to-cancel-police-contracts-and-support-defunding-seattle-pd-e89fa5d9e843

Ghaffary, Shirin. 2020. “The controversy behind a star Google AI researcher’s departure” Vox. https://www.vox.com/recode/2020/12/4/22153786/google-timnit-gebru-ethical-ai-jeff-dean-controversy-fired

Google. 2021. “AI Principles: reviews and operations.” *Google.* https://ai.google/responsibilities/review-process/

Google Walkout for Real Change. 2020. “Standing with Dr. Timnit Gebru — #ISupportTimnit #BelieveBlackWomen.” *Medium.* https://googlewalkout.medium.com/standing-with-dr-timnit-gebru-isupporttimnit-believeblackwomen-6dadc300d382

Googlers Against Transphobia. 2019. “Googlers Against Transphobia and Hate.” *Medium.* https://medium.com/@against.transphobia/googlers-against-transphobia-and-hate-b1b0a5dbf76

Graham, Jill W. 1986. “Principled organizational dissent: A theoretical essay.” *Research in Organizational Behavior*. 8:1-52.

Gray, Mary L. and Siddharth Suri. 2019. *Ghost Work: How to Stop Silicon Valley from Building a New Global Underclass.* New York: Houghton Mifflin Harcourt.

Greene, Daniel, Hoffman, Anna Lauren, and Luke Stark. 2019. “Better, Nicer, Clearer, Fairer: A Critical Assessment of the Movement for Ethical Artificial Intelligence and Machine Learning.” *Proceedings of the 52nd Hawaii International Conference on System Sciences.*

Greene, Jay. 2020. “Amazon fires two tech workers who criticized the company’s warehouse workplace conditions.” The Washington Post. https://www.washingtonpost.com/technology/2020/04/13/amazon-workers-fired/

Hadfield-Menell, Dylan and Gillian K. Hadfield. 2019. *Incomplete Contracting and AI Alignment Proceedings of the 2019 AAAI/ACM Conference on AI, Ethics, and Society.* Presentation. AIES ‘19. New York, NY, USA Honolulu, HI, USA: Association for Computing Machinery.

Hao, Karen. 2020. “We read the paper that forced Timnit Gebru out of Google. Here’s what it says.” MIT Technology Review. https://www.technologyreview.com/2020/12/04/1013294/google-ai-ethics-research-paper-forced-out-timnit-gebru/

Hoofnagle, Chris Jay and Jan Whittington. 2014. "Free: Accounting for the Costs of the Internet's Most Popular Price." *UCLA Law Revie*w. 61:606-669.

Irani, Lilly C. and M. Six Silberman. 2013. *Turkopticon: Interrupting Worker Invisibility in Amazon Mechanical Turk Proceedings of the SIGCHI Conference on Human Factors in Computing Systems.* Presentation. CHI '13. New York, NY, USA Paris, France: Association for Computing Machinery.

Jack, Margaret, and Seyram Avle. 2021. "A Feminist Geopolitics of Technology." *Global Perspectives* 2 (1).

Jobin, Anna, Ienca, Marcello, and Effy Vayena. 2019. "Artificial Intelligence: the global landscape of ethics guidelines." *Nature Machine Intelligence* 1.

Lei, Ya-Wen. 2021. "Delivering Solidarity: Platform Architecture and Collective Contention in China's Platform Economy." *American Sociological Review.* 0003122420979980.

Lynch, Amanda H. and Siri Veland. 2018. *Urgency in the Anthropocene.* Cambridge, MA: The MIT Press.

Maas, Matthijs M. 2018. *Regulating for 'Normal AI Accidents' Operational Lessons for the Responsible Governance of Artificial Intelligence Deployment.* Presentation. Proceedings of the 2018 AAAI/ACM Conference on AI, Ethics, and Society.

McAdam, Doug. 1999. *Political Process and the Development of Black Insurgency, 1930-1970, Second Edition*. Chicago: University of Chicago Press.

McAdam, Doug. 2015. "Collective Action." in *The Wiley Blackwell Encyclopedia of Race, Ethnicity, and Nationalism*. G. Ritzer (Ed.).

Meyer, John W. and Brian Rowan. 1977. "Institutionalized Organizations: Formal Structure as Myth and Ceremony." *American Journal of Sociology.* 83 (2):340-363.

Microsoft. 2021. "Operationalizing responsible AI." *Microsoft.* https://www.microsoft.com/en-us/ai/our-approach

Microsoft Employees. 2019. Microsoft Workers for Climate Justice. *GitHub.* https://github.com/MSworkers/for.ClimateAction

Moss, Emanuel, Watkins, Elizabeth Anne, Singh, Ranjit, Elish, Madeleine Clare and Jacob Metcalf. 2021. "Assembling Accountability: Algorithmic Impact Assessment for Public Interest." *Data & Society.* https://datasociety.net/library/assembling-accountability-algorithmic-impact-assessment-for-the-public-interest/

Noble, Safiya. 2018. *Algorithms of Oppression: How Search Engines Reinforce Racism.* New York: New York University Press.

Parnas, David L., A. John van Schouwen, and Shu Po Kwan. 1990. "Evaluation of Safety-Critical Software." *Commun. ACM.* 33 (6):636-648.

Pasquale, Frank. 2015. *The Black Box Society: The Secret Algorithms that Control Money and Information.* Cambridge, MA: Harvard University Press.

Perrow, Charles. 1984. *Normal Accidents: Living with High-Risk Technologies*. Princeton: Princeton University Press.

Pichai, Sundar. 2018. " AI at Google: our principles." *Google: The Keyword.* https://blog.google/technology/ai/ai-principles/

Posada, Julian. 2020. "The Future of Work Is Here: Toward a Comprehensive Approach to Artificial Intelligence and Labour." *Ethics of AI in Context*. arXiv:2007.05843

Raji, Inioluwa Deborah and Joy Buolamwini. 2019. *Actionable Auditing: Investigating the Impact of Publicly Naming Biased Performance Results of Commercial AI Products Proceedings of the 2019 AAAI/ACM Conference on AI, Ethics, and Society.* Presentation. AIES '19. New York, NY, USA Honolulu, HI, USA: Association for Computing Machinery.

Raji, Inioluwa Deborah, Andrew Smart, Rebecca N. White, Margaret Mitchell, Timnit Gebru, Ben Hutchinson, Jamila Smith-Loud, Daniel Theron, and Parker Barnes. 2020. *Closing the AI Accountability Gap: Defining an End-to-End Framework for Internal Algorithmic Auditing Proceedings of the 2020 Conference on Fairness, Accountability, and Transparency.* Presentation. FAT* '20. New York, NY, USA Barcelona, Spain: Association for Computing Machinery.

Ranganathan, Aruna. 2018. "The Artisan and His Audience:  Identification with Work and Price Setting in a Handicraft Cluster in Southern India." *Administrative Science Quarterly.* 63 (3):637-667.

Renwu Staff. 2020. "Delivery Drivers, Trapped in the System." Renwu Magazine. https://mp.weixin.qq.com/s/Mes1RqIOdp48CMw4pXTwXw

Rosenblat, Alex and Luke Stark. 2016. "Algorithmic Labor and Information Asymmetries: A Case Study of Uber's Drivers." *International Journal of Communication.* 10 (27).

Rosenblat, Alex. 2018. *Uberland: How Algorithms Are Rewriting the Rules of Work.* Oakland: University of California Press.

Russon, Mary-Ann. 2020. "Uber sued by drivers over 'automated robo-firing'." *BBC.* https://www.bbc.co.uk/news/amp/business-54698858

Selbst, Andrew D. 2017. "Disparate impact in big data policing." *Georgia Law Review* 52:109.

Shane, Scott, Metz, Cade, and Daisuke Wakabayashi. 2018. "How a Pentagon Contract Became an Identity Crisis for Google." *The New York Times.* https://www.nytimes.com/2018/05/30/technology/google-project-maven-pentagon.html

Shestakofsky, Benjamin and Shreeharsh Kelkar. 2020. "Making platforms work: relationship labor and the management of publics." *Theory and Society.* 49 (5):863-896.

Si, Ying. 2018. "After experiencing five iterations, Ele.me's "Arc" AI dispatching system controls 3 million drivers." *Aliyun Developer Community Blog.* https://developer.aliyun.com/article/594688

Tan, JS and Nataliya Nedzhvetskaya. (2020). *Collective Action In Tech.* https://data.collectiveaction.tech/

Tarnoff, Ben. 2020. "The Making of the Tech Worker Movement." *LOGIC.*

Tech Won't Drill It. 2020. *Tech Won't Drill It.* Medium.
https://medium.com/@techwontdrillit/tech-wont-drill-it-a63594dc6e66

Tiku, Nitasha. 2019. "Google Walkout Organizers Say They're Facing Retaliation." *WIRED.*
https://www.wired.com/story/google-walkout-organizers-say-theyre-facing-retaliation/

Useem, Michael. 1993. *Executive Defense: Shareholder Power and Corporate Reorganization*. Cambridge: Harvard University Press.

Vaughan, Diane. 1996. *The Challenger Launch Decision: Risky Technology, Culture and Deviance at NASA*. Chicago.

Wachter, Jan K. and Patrick L. Yorio. 2014. "A system of safety management practices and worker engagement for reducing and preventing accidents: An empirical and theoretical investigation." *Accident Analysis & Prevention*. 68: 117-130.

Wakabayashi, Daisuke and Katie Benner. 2018. "How Google Protected Andy Rubin, the 'Father of Android'" New York Times.
https://www.nytimes.com/2018/10/25/technology/google-sexual-harassment-andy-rubin.html

Whittlestone, Jess, Nyrup, Rune, Alexandrova, Stephen Cave. 2019. *The role and limits of principles in AI ethics: towards a focus on tensions.* Presentation. Proceedings of the 2019 AAAI/ACM Conference on AI, Ethics, and Society.

Williams, Lauren. 2020. "JEDI delay slows DOD's AI push" *Defense Systems.*
https://defensesystems.com/articles/2020/05/27/jedi-delay-slows-ai.aspx